# Stress-Induced Mutagenesis and Complex Adaptation

Yoav Ram and Lilach Hadany

**Email**: YR - yoavram@post.tau.ac.il,

LH - lilach.hadany@gmail.com (corresponding author)

**Address**: Dept. of Molecular Biology and Ecology of Plants, Tel Aviv University

Tel-Aviv 69978, Israel. Tel. +972.3.640.6886

## Summary

Because mutations are mostly deleterious, mutation rates should be reduced by natural selection. However, mutations also provide the raw material for adaptation. Therefore, evolutionary theory suggests that the mutation rate must balance between *adaptability* – the ability to adapt – and *adaptedness* – the ability to remain adapted. We model an asexual population crossing a fitness valley and analyze the rate of complex adaptation with and without stress-induced mutagenesis – the increase of mutation rates in response to stress or maladaptation. We show that stress-induced mutagenesis increases the rate of complex adaptation without reducing the population mean fitness, thus breaking the evolutionary trade-off between *adaptability* and *adaptedness*. Our theoretical results support the hypothesis that stress-induced mutagenesis promotes adaptation and provide quantitative predictions of the rate of complex adaptation with different mutational strategies.

## 1. Introduction

There is experimental, clinical and theoretical evidence that high mutation rates increase the rate of adaptation and that during adaptive evolution, constitutive mutators - alleles that constitutively increase the mutation rate - can rise in frequency because of the beneficial mutations they generate [1–3]. However, during evolution in a stable environment, constitutive mutators become associated with poor genetic backgrounds due to increased accumulation of deleterious mutations; this was evidenced both in the lab [4] and in the clinic [5]. Classical models suggest the "reduction principle", which states that natural selection reduces the mutation rate in a stable environment [6,7]. But many adaptations require new beneficial mutations, especially in asexual populations. This tension between the effects of beneficial and deleterious mutations leads to "the rise and fall of the mutator allele" [8], where mutator alleles increase in frequency in a maladapted population, only to be eliminated by natural selection when the population is well-adapted. This dynamic was studied using experimental evolution [9,10], mathematical analysis, and simulations [11–13].

Thus, the mutation rate must balance between two evolutionary traits, as Leigh [14] suggested: *adaptability* – the capacity to adapt to new environmental conditions – and *adaptedness* – the capacity to remain adapted to existing conditions.

Stress-induced mutagenesis (SIM) - the increase of mutation rates in stressed or maladapted individuals - has been demonstrated in several species, including both prokaryotes and eukaryotes [15]. SIM has been observed in lab strains [16,17] and natural populations of *Escherichia coli* [18, but see 19], and in other species of bacteria such as Pseudomonads [20], *Helicobacter pylori* [21], *Vibrio cholera* [22] and *Streptococcus pneumonia* [23]. SIM has also been observed in yeast [24,25], algae [26], nematodes [27], flies [28], and human cancer cells [29]. Several stress responses regulate the mutation rate in bacteria by shifting replication to error-prone DNA polymerases [30] and by inhibiting the mismatch repair system [31]. These stress responses include the SOS DNA-damage response, the RpoS-controlled general or starvation stress response, and the RpoE membrane protein stress response [32].

It is still not clear how SIM affects evolution and adaptation. Some authors have proposed that SIM has a significant impact on *adaptability* or *evolvability* [17,33,34], but there is no theoretical treatment of this impact. On the other hand, the effect of SIM on *adaptedness* was studied with deterministic [35] and stochastic [36] models. These articles showed that without beneficial mutations SIM doesn't affect the mean fitness of asexual populations in stable environments, in contrast with constitutive mutagenesis, which decreases the population mean fitness. More recently, we have shown that with rare beneficial mutations, if maladapted individuals increase their mutation rate then the population mean fitness of asexual populations increases [37].

Here, we analyze population genetics models of adaptive evolution to explore the rate of complex adaptation on rugged fitness landscapes, in which adaptations require two separately deleterious mutations [38,39]. We develop analytic approximations and stochastic simulations and compare normal, constitutive, and stress-induced mutagenesis. We show that stress-induced mutagenesis can break the trade-off between *adaptability* and *adaptedness* by increasing the rate of complex adaptation without decreasing the population mean fitness.

## 2. Model

We model a population of *N* haploid asexual individuals with a large number of loci in full linkage. The model includes the effects of mutation, selection, and genetic drift. Individuals are characterized by their genotype in two specific bi-allelic loci – *ab*, *Ab*, *aB*, and *AB* – and by the number of deleterious mutations they carry in the rest of the non-specific loci. For example, *aB/3* is the *aB* genotype with additional three deleterious mutations in non-specific loci.

We focus on adaptation to a new rugged fitness landscape. The fitness of the wildtype *ab/0* is 1, the fitness of the single mutants *Ab/0* and *aB/0* is 1-*s*, and the double mutant *AB/0* has the highest fitness 1+*sH*, where *s* is the selection coefficient and *H* is the relative advantage of the double mutant. This is the simplest case of a rugged fitness landscape: the single mutants *Ab* and *aB* are fitness valleys between the local and global fitness peaks *ab/0* and *AB/0* (Figure 1).

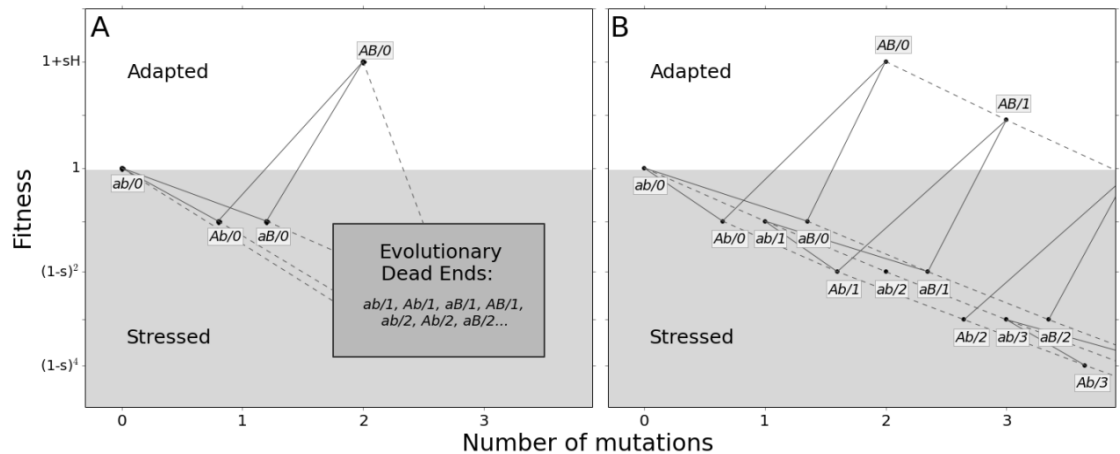

**Figure 1 – Adaptation on a rugged fitness landscape.** The figure shows the fitness of the possible genotypes, which are represented by the allele combination at the specific loci (*ab*, *Ab*, *aB*, and *AB*) and the number of deleterious alleles across the genome following the forward-slash ('/'). The y-axis represents fitness: the wildtype *ab*/0 has fitness 1; the fittest genotype *AB*/0 has fitness $1+sH$; deleterious alleles, either at the *A/a* and *B/b* loci, or at the non-specific loci, reduce fitness by $1-s$. The x-axis represents the number of accumulated mutations (genotypes are jittered to increase separation). Solid lines represent mutations at the *a/A* and *b/B* loci, occurring with probability $\mu$. Dashed lines represent deleterious mutations in the rest of the genome, occurring with rate $U$. Mutagenesis is induced in stressed genotypes with fitness <1 (gray background). Fit genotypes, with fitness ≥1, do not hypermutate (white background). **(A)** In the analytic model genotypes with deleterious alleles in non-specific loci are considered "Evolutionary Dead Ends" and do not contribute to adaptation. **(B)** In the simulations individuals can accumulate up to 25 deleterious alleles (the figure only shows three). Multiple mutations can occur simultaneously but are not shown for simplicity of the illustration.

Deleterious mutations in the non-specific loci independently (multiplicatively) reduce the fitness of the individual by $1-s$. Mutations occur in the specific loci with probability $\mu$. The number of new mutations per replication in the rest of the genome (the non-specific loci) is Poisson distributed with an average $U$. The model neglects back-mutations and compensatory mutations due of their minor short term effects.

We consider three mutational strategies: normal mutagenesis (NM), where there is no increase in the mutation rate; constitutive mutagenesis (CM), where all individuals always increase their mutation rate by $\tau$, the mutation rate fold increase;

and stress-induced mutagenesis (SIM), where only stressed or maladapted individuals increase their mutation rate by $\tau$. Individuals are considered stressed if their fitness is below a specific threshold, so stress can be caused by a deleterious mutation (either in the specific *A/a* and *B/b* loci or in non-specific loci). The main analysis assumes that the effect of SIM on the mutation rate of an individual with fitness $\omega$ is

$$U(\omega) = \begin{cases} \tau U, & \omega < 1 \\ U, & \omega \geq 1 \end{cases}. \tag{1}$$

This equation models a scenario in which an environmental change – *i.e.*, appearance of a new ecological niche or a new carbon source – provides an opportunity for adaptation without affecting the fitness of the wildtype (*ab/0*). We also study a different scenario in which the environmental change reduces the absolute fitness of the wildtype so that it is also stressed – see section 3.5.

We are interested in calculating the adaptation rate of a population homogenous for each of the above mutational strategies (NM, CM, or SIM). The adaptation process is separated into two distinct stages. In the first stage, a double mutant *AB* appears in the population, usually in a single copy. In the second stage, the double mutant either goes to extinction or avoids extinction, increases in frequency, and goes to fixation.

We analyzed this model with two methods. The first is analytic (Figure 1A), in which we assume that: (i) genotypes with deleterious backgrounds (deleterious alleles in the non-specific loci) do not contribute to the adaptation process; and (ii) the number of deleterious alleles per individual before the appearance of a double mutant is at a mutation-selection balance (MSB) and is Poisson distributed with mean $U/s$ (Haigh 1978). The former assumption requires that mutation is weaker than selection ($U \ll s$); the later assumption only requires that mutation is not much stronger than selection. Specifically, the expected number of mutation-free individuals is at least one: $Ne^{-U/s} > 1 \Rightarrow U < s \cdot logN$ [41].

The second method is a stochastic Wright-Fisher simulation with selection, mutation and genetic drift (Figure 1B), in which: (i) individuals with a deleterious background can contribute to adaptation; (ii) a mutation-free population evolves towards a MSB without assuming a Poisson distribution of the number of deleterious alleles.

## 2.1. Wright-Fisher simulations

We track the number of individuals in each genotype class: *ab*/*x*, *Ab*/*x*, *aB*/*x*, and *AB*/*x*, where *x*≥0 is the number of deleterious alleles in non-specific loci. The simulations start with a single-peak smooth fitness landscape (the fitness of *AB*/*x* is $(1-s)^{2+x}$) and a mutation-free population (all individuals start in the optimal *ab*/*0* genotype with fitness 1) that accumulates deleterious mutations over the first 5,000 generations of the simulation. With *s*=0.05 and 0.005, 180 and 1,800 generations are enough for the average number of deleterious alleles per individual to reach 99.99% of its MSB value, $U/s$ [42].

After 5,000 generations the fitness landscape changes to a rugged one, making *AB* the optimal genotype with fitness $1+sH$ (Figure 1B). The simulation then proceeds until an *AB* genotype appears and either fixates in the population or goes extinct (either all or no individuals are in the *AB* classes, respectively). Therefore, each simulation provides one sample of the waiting time for the appearance of a double mutant and one sample of the probability of fixation of a double mutant. At least 1,000 simulations were performed for each parameter set.

Table 1 summarizes the model parameters with estimated values for *E. coli*.

Table 1 – Model parameters and estimated values for *E. coli*

| Symbol | Name | Estimate | References |
|---|---|---|---|
| $s$ | Selection coefficient | 0.001-0.03 | [43,44] |
| $H$ | Double mutant advantage | 1-10 | [44] |
| $U$ | Genomic deleterious mutation rate | 0.0004-0.003 | [45,46] |
| $\mu$ | Site-specific mutation rate | $U$/5000 | [44] |
| $\tau$ | Fold-increase in mutation rate | 1-100 | [18,47] |
| $N$ | Population size | $10^5$-$10^{10}$ | [48,49] |

# 3. Results

## 3.1. Appearance of a double mutant

We are interested in the waiting time for the appearance of a double mutant *AB* either by a double mutation in a wildtype individual *ab*, or via a single mutation in a single mutant *Ab* or *aB* (Figure 1A). Denoting the population size by *N*, we note that (i) if $Ne^{-U/s}(\mu/s)^2 > 1$ then double mutants are already expected at the MSB and adaptation will not require new mutations; (ii) if $Ne^{-U/s}\mu/s < 1$ then no single mutants are expected at the MSB and double mutants must be generated by a double mutation in a wildtype individual. In this case, increasing the mutation rate of individuals with fitness below 1 will have no effect on the appearance of the double mutant and there is no point in analyzing the effect of SIM.

Combining the two constraints we get this constraint on the population size *N*: $e^{U/s} s/\mu < N < e^{U/s}(s/\mu)^2$. This constraint is reasonable for bacterial populations (see Table 1).

The frequencies of wildtype (*ab*) and single mutants (*aB* and *Ab* combined) that are mutation-free at the MSB are roughly $e^{-U/s}$ and $2\mu/s \cdot e^{-U/s}$, respectively. The probability that an offspring of a wildtype or single mutant parent is a double mutant *AB* is $\mu^2$ and $\mu$, respectively. The probability that such an offspring is also mutation-free in the rest of its genome (the only mutations that occurred were at the specific loci) is $e^{-U}$. Therefore, the probability $q$ that a random offspring is a double mutant, given there are no double mutants in the current generation, is approximated by

$$q = \mu^2 e^{-\frac{U}{s}-U} + 2\frac{\mu^2}{s}e^{-\frac{U}{s}-U} \approx 2\frac{\mu^2}{s}\left(1-\frac{U}{s}\right). \tag{2}$$

The first expression assumes that individuals with a deleterious background don't contribute to adaptation and that the MSB distribution of deleterious alleles is Poisson. The second expression also assumes that mutation is much weaker than selection: *U*<<s.

With SIM the mutation rate of single mutants is increased $\tau$-fold and the probability that a random offspring is a double mutant is

$$q_{SIM} = \mu^2 e^{-\frac{U}{s}-U} + 2\frac{\tau\mu^2}{s} e^{-\frac{U}{s}-\tau U} \approx q \cdot \tau(1-\tau U). \tag{3}$$

These expressions use the same assumptions as in eq. 2. The second expression also assumes that $\tau U < 1$.

Appendix A includes full derivations of the above equations and Figure S1 compares them with simulations results.

## 3.2. Fixation probability of the double mutant

Assuming an advantage to the double mutant ($H>1$) and a large population size (see the above constraint on $N$), a double mutant has two possible fates after its appearance: fixation or extinction. Following Eshel [50], the fixation probability $\rho$ of the double mutant is (see Appendix B)

$$\rho \approx 2\frac{sH}{1+sH} \approx 2sH. \tag{4}$$

That is, the fixation probability of the double mutant is roughly twice its adaptive advantage. This is a classic result of population genetics theory [51,52].

The fixation probability with SIM equals that of NM and CM because the mutation rate of the wildtype *ab* equals that of the double mutant *AB* (but see an exception in section 3.5).

## 3.3. Adaptation rate

From the probability $q$ that a random offspring is a double mutant, we can derive the probability that one or more double mutants appear in the next generation: $1-(1-q)^N \approx Nq$. This is a good approximation because $Nq$ is very small due to the constraint on $N$. Once a double mutant appears it goes to fixation with probability $\rho$.

When fixation is much faster than appearance of the double mutant *AB*, the time for adaptation *T* can be approximated by the waiting time for a double mutant that goes to fixation. This waiting time follows a geometric distribution with rate $Nq\rho$ and therefore the adaptation rate *v* (the inverse of the waiting time for adaptation) is approximately

$$v = E[T]^{-1} \approx Nq\rho. \tag{5}$$

Plugging eqs. 2-4 in eq. 5, we get these approximations:

$$v_{NM} = 2NH\mu^2 e^{-\frac{U}{s}-U}(2+s) \approx 4NH\mu^2\left(1-\frac{U}{s}\right) \tag{6}$$

$$v_{CM} = v_{NM} \cdot \tau^2 e^{\frac{-(\tau-1)U(1+s)}{s}} \approx v_{NM} \cdot \tau^2\left(1-\frac{\tau U}{s}\right) \tag{7}$$

$$v_{SIM} = v_{NM} \cdot \frac{2\tau e^{-(\tau-1)U}+s}{2+s} \approx v_{NM}\cdot \tau(1-\tau U) \tag{8}$$

NM is normal mutagenesis, CM is constitutive mutagenesis, and SIM is stress-induced mutagenesis. The middle expression in each equation is the full approximation, which assumes a Poison distribution and no contribution of deleterious genotypes to adaptation. The right hand sides are first order approximations that assume mutation is much weaker than selection ($U \ll s$ for NM and SIM, $\tau U \ll s$ for CM) and that $1<\tau<1/U$. See Table 1 for description of model parameters and an article by Weinreich and Chao [53] for a result similar to eq. 6.

The main conclusions from eqs. 6-8: First, adaptation with CM is faster than with NM. Second, adaptation with SIM is also faster than with NM, but not as fast as with CM because the mutation-free wildtype (*ab/0*) does not hypermutate.

If mutation is weaker than selection ($U \ll s$) then the adaptation rate with CM increases with $\tau^2$ and the adaptation rate with SIM increases with $\tau$. In addition, because the fixation probability is the same for NM, CM and SIM, the differences in the adaptation rate are due to differences in the appearance probability $q$ (Figure S1); see section 3.5 for a different scenario in which SIM also increases the fixation probability.

Figure 2 compares the analytic approximations with simulation results for the weak mutation regime ($U \ll s$). This regime is relevant for asexual microbes in which the deleterious mutation rate is generally $10^{-4}$-$10^{-3}$ mutations per genome per generation and selection coefficients are estimated to be between $10^{-1}$ and $10^{-2}$ (see Table 1). When the mutation rate fold increase $\tau$ is high (>10), the approximations slightly overestimate the adaptation rate because the double mutant *AB* is more likely to appear on a deleterious background (*AB/1* instead of *AB/0*). Because the fitness of *AB/1* is higher than that of the wildtype *ab/0* (this happens because $H>(1-s)^{-1}\approx 1+s$), the double mutant can go to fixation even when it appears on a deleterious background, sweeping the deleterious alleles with it to fixation in a process called "genetic hitch-hiking" [54]. However, these sweeps result in a lower fixation probability for the double mutant (Figure S2).

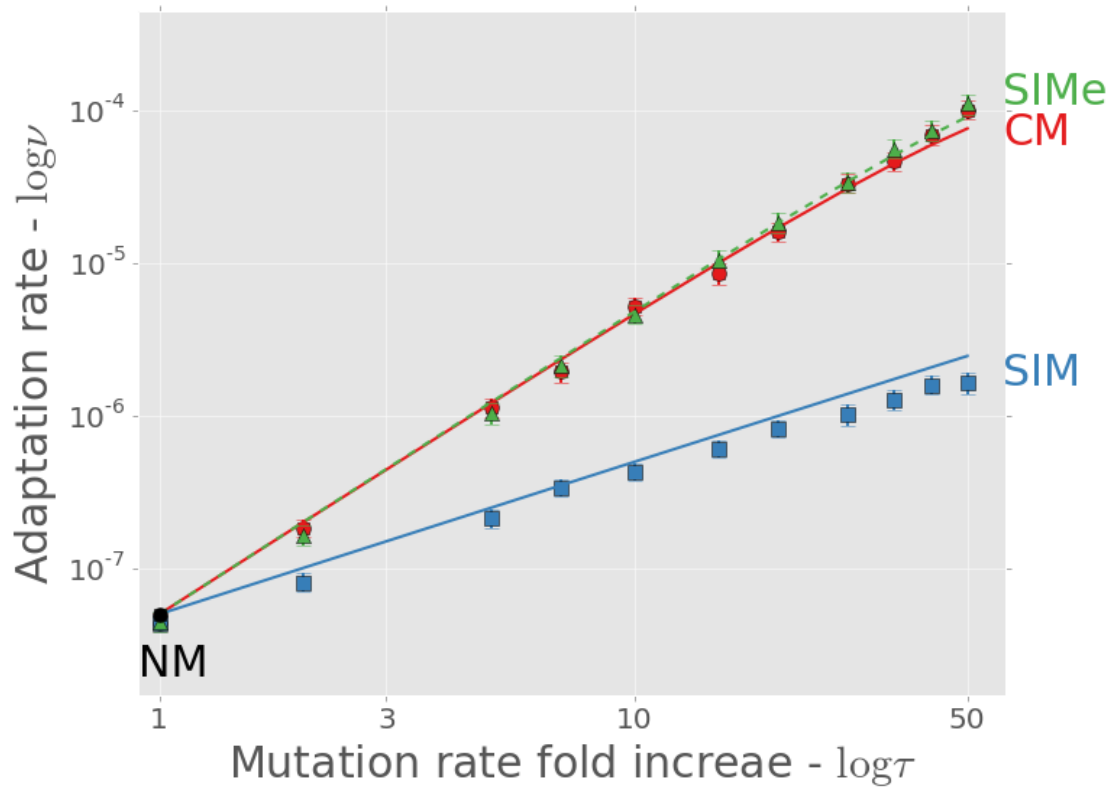

**Figure 2 – Complex adaptation with different mutational strategies.** The figure shows the adaptation rate $v$ as a function of the mutation rate increase $\tau$ (both in log scale). A black circle is normal mutagenesis (NM; $\tau=1$); solid line with circles is constitutive mutagenesis (CM); solid line with squares is stress-induced mutagenesis (SIM); dashed lines with triangles is stress-induced mutagenesis with environmental stress (SIMe; see section 3.5). Lines are analytic approximations. Markers are the means of stochastic simulation results. Error bars represent 95% confidence interval of the mean (at least 1,000 simulations per point; computed with bootstrap with 1,000 samples per point). Parameters (see Table 1): $U=0.0004$, $s=0.05$, $\beta=0.0002$, $H=2$, $N=10^6$.

What happens when mutation is as strong as selection? Figure 3A shows results for $s=10U$. When the average number of deleterious alleles per individual $\tau U/s$ is over one, adaptation with CM is likely to occur on a deleterious background. Because our approximation neglects adaptation on deleterious backgrounds, it underestimates the adaptation rate (Figure 3A). Note that although the adaptation rate continues to increase with $\tau$, the population carries more deleterious alleles after adaptation, resulting in a lower population mean fitness (Figure 3B) and eventually a lower fixation probability and adaptation rate (Figure 3A) .

With SIM, the average number of deleterious alleles per individual $U/s$ does not increase with $\tau$, because mutation-free individuals (*ab/0*) do not hypermutate. As in the case of weak mutation, when $\tau>10$ the double mutant can appear on a deleterious background, resulting in hitch-hiking and a lower fixation probability, and causing our approximation to overestimate the adaptation rate (Figure 3A).

### 3.4.    The trade-off between *adaptability* and *adaptedness*

Next, we explore how different mutational strategies (NM, CM and SIM) balance between *adaptability* – the ability to adapt to new conditions – and *adaptedness* – the ability to remain adapted to current conditions. For this purpose we define *adaptedness* as $\bar{\omega}$ the population mean fitness in a stable environment and *adaptability* as $v$ *the* rate of complex adaptation.

We used the above approximations (eqs. 6-8) to calculate the rate of complex adaptation of populations with NM, CM and SIM. We also extended an existing model [37] to calculate the population mean fitness at the mutation-selection balance.

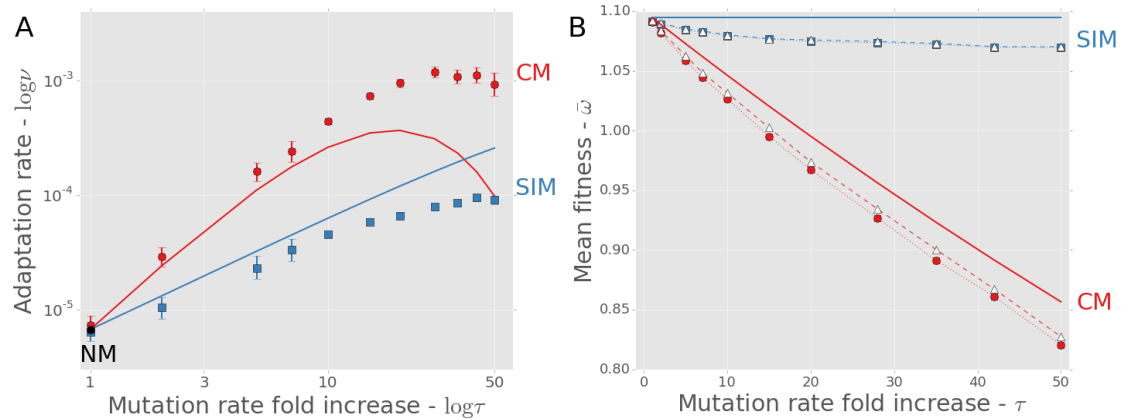

**Figure 3 – Adaptation with strong mutation.** When the deleterious mutation rate is high – here $U=s/10$ – the adaptation process can lead to hitch-hiking of deleterious alleles with the beneficial double mutant. **(A)** The adaptation rate $v$ as a function of the mutation rate increase $\tau$ (both in log scale). A black circle for normal mutagenesis (NM; $\tau=1$); red solid line and circles for constitutive mutagenesis (CM); blue solid line and squares for stress-induced mutagenesis (SIM). Lines are analytic approximations. Markers are the means of stochastic simulations results. Error bars represent 95% confidence interval of the mean (at least 1,000 simulations per point; computed with bootstrap with 1,000 samples per point). Parameters (see Table 1): $U=0.005$, $s=0.05$, $\beta=0.0002$, $H=2$, $N=10^6$. **(B)** The population mean fitness $\bar{\omega}$ after successful fixation of the beneficial double mutant as a function of the mutation rate increase $\tau$. Solid lines are analytic approximations neglecting adaptation from deleterious background ($e^{-\tau U}(1+sH)$); dotted lines with filled squares (SIM) and circles (CM) are the means of stochastic simulation results; dashed lines with white triangles are predictions based on the genotype on which $AB$ appeared in the simulations, including MSB but disregarding the effects of drift during the fixation process (which only has a significant effect with CM due to higher mutation rates in the wildtype). Error bars are too small to see. Same parameter values as in panel A.

This extended model includes rare back- or compensatory mutations (which have a stronger effect on mutation-selection balance dynamics than on adaptive dynamics) and allows more than one mutation to occur in the same individual and generation. The details of this model and the calculation of population mean fitness with various mutational strategies are given in Appendix D.

The mutation rate with CM is constant and uniform across the population, and the population mean fitness mainly depends on the fitness and mutation rate of the fittest individuals. Therefore, the population mean fitness decreases when the mutation rate increases; this decrease is due to generation of deleterious mutations in the fittest individuals. The adaptation rate, however, increases with the mutation rate (eq. 7). This trade-off between *adaptability* and *adaptedness* constraints the population: after a long period of environmental stability it can lose the potential for adaptation, and after a long period of environmental change the population can be susceptible to reduced fitness and mutational meltdowns [55].

However, this trade-off between *adaptability* and *adaptedness* can be broken if mutation rates are not uniform across the population. Increased mutation rates in unfit individuals increase the population mean fitness, as long as beneficial (or compensatory) mutations can occur [37]. Figure D1 shows this advantage of SIM over NM in terms of the difference in population mean fitness ($\bar{\omega}_{SIM} - \bar{\omega}_{NM}$). Moreover, increased mutation rates in unfit individuals also increase the adaptation rate (eq. 8; Figure 2). Therefore, SIM breaks the trade-off between *adaptability* and *adaptedness*.

Figure 4 shows the adaptation rate and population mean fitness of CM and SIM compared to NM for different values of $\tau$, the mutation rate fold increase.

Any realistic rate of adaptation *v* can be realized using both CM and SIM. The highest mean fitness will always be attained with SIM, which has a small advantage over NM (that cannot be seen in this figure, but see Figure D1) due to the increased generation of beneficial mutations in individuals with low fitness. If for some rate of adaptation the mutation rate fold increase $\tau$ required by SIM is too high (*i.e.*, *τU>s*), the same adaptation rate can be realized by a mixed strategy (dashed line in Figure 4). For example, a 96-fold increase in adaptation rate can be achieved with CM with $\tau=10$, with SIM with $\tau=96$, or with a mixed strategy with $\tau_{CM}=7$ and $\tau_{SIM}=2$ in which all individuals increase their mutation rate 7-fold and stressed individuals further increase their mutation rate 2-fold. However, these increases in adaptation rates have a price: the mutational load will decrease the population mean fitness from 0.9996 with NM to 0.996 with CM and 0.9972 with the mixed strategy. This price in not paid by populations with SIM because the mean fitness mainly depends on the mutation rate of fit individuals.

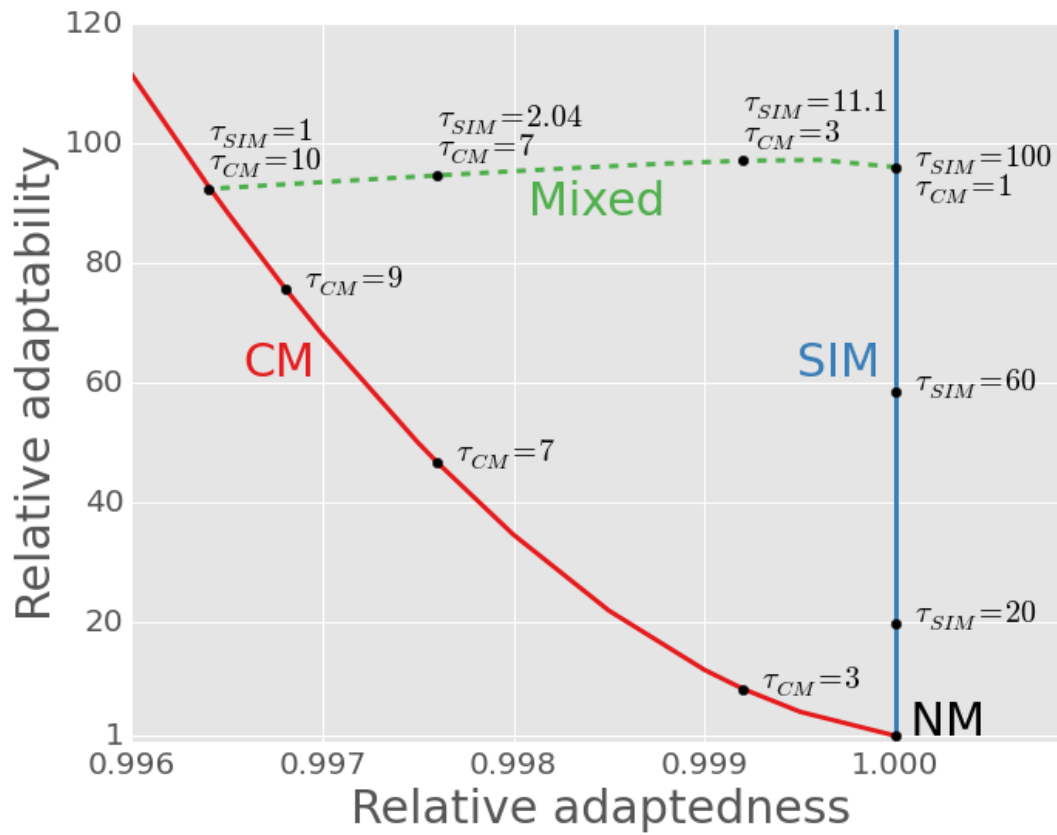

**Figure 4 – The trade-off between *adaptedness* and *adaptability*.** The figure shows the relative *adaptedness* and the relative *adaptability* of different mutational strategies in comparison to normal mutagenesis (NM). *Adaptedness* is defined by the population mean fitness at MSB, $\bar{\omega}$ (see Appendix D). *Adaptability* is defined by the rate of complex adaptation, $v$ (eqs. 6-8). Constitutive mutagenesis (CM) increases the mutation rate of all individuals $\tau_{CM}$-fold; Stress-induced mutagenesis (SIM) increases the mutation rate of stressed individuals $\tau_{SIM}$-fold; Mixed strategies (dashed line) increase the mutation rate of all individuals $\tau_{CM}$-fold and of stressed individuals an additional $\tau_{SIM}$-fold. SIM breaks off the *adaptability-adaptedness* trade-off of CM, increasing the *adaptability* without compromising the *adaptedness* of the population. Parameters (see Table 1): $N=10^6$, $U=0.0004$, $\beta=0.0002$, $s=0.05$, $H=2$, $\tau \ll s/U$.

### 3.5. Environmental stress

So far, we considered the case where the environmental change creates an opportunity for adaptation without affecting the absolute fitness of the population – for example, a new ecological niche can be favorable without affecting the well-being of the current population. In that scenario, the wildtype *ab* was not stressed and did not hypermutate.

Next, we consider a different scenario in which an environmental change affects the well-being of the entire population: for example, exposure to an antibiotic drug or a host immune response. In this case the environmental change doesn't just create an opportunity for adaptation but also causes stress in the entire population. We use a subscript *e* to denote quantities related with this scenario.

As before the double mutant *AB* is resistant to the stress (*i.e.* the drug or immune response) and therefore has a higher fitness than either the wildtype or the non-resistant single mutant*s*. However, in this scenario the wildtype *ab* is also stressed and therefore hypermutates with SIM (compare with eq. 1):

$$U_e(\omega) = \begin{cases} U, & \omega > 1 \\ \tau U, & \omega \leq 1 \end{cases}. \tag{9}$$

This scenario has an important biological relevance, as SIM has been implicated in the evolution of drug resistance in bacteria and yeast [34,56,57] and could be involved in the evolution of pathogen virulence and the evolution of drug resistance and progression in cancer cells [58].

We assume that after the environmental change the SIM$_e$ population has reached a new MSB [42] with mutation rate $\tau U$, before the appearance of the double mutant (with $s$=0.05 and $U$=0.0004, for example, the average number of deleterious mutations is 0.99·$U/s$ after 90 generations, whereas the adaptation time is well over 1,000 generations). Under this assumption, the adaptation rate with SIM$_e$ is (see Appendix C for full derivation)

$$v_{SIM_e} \approx v_{CM} \cdot \left(1 + \frac{U(\tau-1)}{sH}\right). \tag{10}$$

That is, adaptation with SIM$_e$ is faster than with CM (Figure 2A). The fixation probability of double mutants is higher with SIM$_e$ than with CM, because the mutation rate of double mutants is lower than that of the rest of the population. This difference in mutation rates confers an additional selective advantage to the double mutants (see Appendix C) which increases their fixation probability:

$$\rho_{SIM_e} \approx \rho \left(1 + \frac{U(\tau-1)}{sH}\right). \tag{11}$$

This additive advantage increases linearly with $\tau$ with a slope of $U/sH$ and can be significant: for $s$=0.05, $H$=2 and $U$=0.0004, increasing the mutation rate of stressed individuals 10-fold increases the fixation probability by 3.6%. The increased fixation probability was verified by simulations (Figure S2).

## 4. Discussion

We studied the effect of stress-induced mutagenesis (SIM) on both the *adaptability* – the capacity of populations to adapt to new complex conditions– and the *adaptedness* – the ability of populations to stay adapted to existing conditions [14]. We showed that SIM breaks the trade-off between *adaptability* and *adaptedness*, allowing rapid adaptation to complex environmental challenges without compromising the population mean fitness in a stable environment.

In addition to the pure strategies of constitutive mutagenesis (CM) and SIM, our model also considers a mixed mutational strategy. There are two examples of such a mixed strategy. First, if individuals have incomplete information regarding their condition (this is the case in most realistic biological scenarios) then we expect errors in the induction of mutagenesis: induction of mutagenesis without stress and failure to induce mutagenesis under stress. In this case the population would, on average, use a mixed strategy. Second, a mutator allele can increase the mutation rate constitutively and further increase it under stress – for example, a recent study with *Pseudomonas aeruginosa* found that although the *mutS*, *mutY* and *mutM* mutator alleles always increase the mutation rate in comparison with the wildtype, the level of this increase depends on the level of stress the cell experiences [59].

Our model does not assume direct fitness costs for any of the mutational strategies. A "cost of DNA replication fidelity" [60] – the energy and time expended in order to maintain a low mutation rate – could make both CM and SIM more successful. The "cost of fidelity" may require further study, but empirical evidence suggests that it doesn't play an important role in the evolution of the mutation rate [61–64]. Another fitness cost might be associated with the regulation of the mutation rate: for individuals to determine if their condition calls for the induction of mutagenesis, they must invest resources and energy in costly sensory mechanisms. However, such mechanisms already exist for various unrelated purposes, such as the maintenance of cell cycle and homeostasis. Therefore, we consider these mechanisms as "free" in terms of fitness costs. Moreover, in *E. coli* mutagenesis is induced by several stress responses that serve other cellular functions [16,32], and this is probably the case in other organisms as well.

Our model focuses on asexual populations, ignoring recombination, segregation, and sexual reproduction. These mechanisms are important for adaptation on a rugged fitness landscape both because they help to cope with deleterious mutations and because they allow different single mutants to produce double mutants without an increased mutation rate. We expect that recombination will reduce the advantage of SIM over NM in terms of population mean fitness [35], direct competitions [65], and adaptation rate (due to the Fisher-Muller effect).

Mean fitness and adaptation rate are both population-level traits. But simply because SIM has the most efficient balance between these traits doesn't mean it will necessarily evolve, because individual-level selection and population-level selection can act in opposing directions. In a previous article we have demonstrated that $2^{nd}$ order selection can lead to the evolution of SIM [37]: in an asexual population evolving on a smooth fitness landscape, selection favored SIM over both NM and CM. In the current article we show that selection also favors SIM on a rugged fitness landscape (Appendix F).

Complex traits, coded by multiple genes, present an open evolutionary problem, first described by Sewall Wright in 1931: if different alleles are separately deleterious but jointly advantageous, how can a population evolve from one co-adapted gene complex to a fitter one, crossing a less fit "valley"? Wright suggested the "shifting-balance theory of evolution" [38,39]. His solution is valid [66–68] but possibly limited to specific parameter ranges [69–72]. As a result, other mechanisms have been proposed: increased phenotypic variance after population bottlenecks [73]; environmental fluctuations [74]; environmental heterogeneity [75]; fitness-associated recombination [76]; stochastic tunneling in large asexual populations [77]; and intermediate recombination rates [78]. Our model of complex adaptation is similar to that of Weinreich and Chao [53], but our model includes various mutational strategies and the effects of stress and deleterious mutations. Our results (Figure 2) suggest that SIM can help resolve the problem of fitness valley crossing by reducing the time required for a population to shift an adaptive peak.

Our results provide theoretical basis to the conjecture that SIM facilitates adaptation. This conjecture can be tested experimentally, for example, with *E. coli*, where it is possible to interfere with the regulation of mutagenesis [34]. The adaptation time with and without SIM can be measured in an experimental population adapting on a two-peak fitness landscape [79]. These measurements can then be compared to our analytic approximations to determine the relative advantage and disadvantage of the different mutational strategies.

## 4.1. Conclusions

Stress-induced mutagenesis has been implicated as a driver of adaptive evolution for several decades [17,33,80]. We provide theoretical treatment of this concept. Our results show that stress-induced mutagenesis increases the rate of complex adaptation, and that in contrast to constitutive mutagenesis it does not jeopardize the fitness of populations under stable conditions. Because mutation is a fundamental force in every biological system, these results have important implications on many fields in the medical and life sciences, including epidemiology, oncology, ecology, and evolutionary biology.

# 5. Acknowledgments

We thank U. Obolski for advice on statistical analysis. We are grateful to A. F. Agrawal helpful comments on an earlier version of the manuscript. This research has been supported in part by the Israeli Science Foundation 1568/13 (LH) and by Marie Curie reintegration grant 2007–224866 (LH).

Ram & Hadany                                    SIM and Complex AdaptationRam & Hadany                                    SIM and Complex Adaptation

29

50. Eshel, I. 1981 On the survival probability of a slightly advantageous mutant gene with a general distribution of progeny size - a branching process model. *J. Math. Biol.* **12**, 355–362. (doi:10.1007/BF00276922)

51. Fisher, R. A. 1930 *The Genetical Theory of Natural Selection*. Oxford: Clarendon Press.

52. Patwa, Z. & Wahl, L. M. 2008 The fixation probability of beneficial mutations. *J. R. Soc. Interface* **5**, 1279–89. (doi:10.1098/rsif.2008.0248)

53. Weinreich, D. M. & Chao, L. 2005 Rapid evolutionary escape by large populations from local fitness peaks is likely in nature. *Evolution* **59**, 1175–82.

54. Maynard Smith, J. & Haigh, J. 1974 The hitch-hiking effect of a favourable gene. *Genet. Res.* **23**, 23–35. (doi:10.1017/S0016672300014634)

55. Lynch, M., Bürger, R., Butcher, D. & Gabriel, W. 1993 The mutational meltdown in asexual populations. *J. Hered.* **84**, 339–44.

56. Obolski, U. & Hadany, L. 2012 Implications of stress-induced genetic variation for minimizing multidrug resistance in bacteria. *BMC Med.* **10**, 1–30. (doi:10.1186/1741-7015-10-89)

57. Shor, E., Fox, C. a. & Broach, J. R. 2013 The yeast environmental stress response regulates mutagenesis induced by proteotoxic stress. *PLoS Genet.* **9**, e1003680. (doi:10.1371/journal.pgen.1003680)

58. Karpinets, T., Greenwood, D., Pogribny, I. & Samatova, N. 2006 Bacterial stationary-state mutagenesis and Mammalian tumorigenesis as stress-induced cellular adaptations and the role of epigenetics. *Curr. Genomics* **7**, 481–96.

59. Torres-Barceló, C., Cabot, G., Oliver, A., Buckling, A. & MacLean, R. C. 2013 A trade-off between oxidative stress resistance and DNA repair plays a role in the evolution of elevated mutation rates in bacteria. *Proc. R. Soc. B Biol. Sci.* **280**, 20130007. (doi:10.1098/rspb.2013.0007)

60. Dawson, K. J. 1998 Evolutionarily stable mutation rates. *J. Theor. Biol.* **194**, 143–57. (doi:10.1006/jtbi.1998.0752)

61. Giraud, A., Matic, I., Tenaillon, O., Clara, A., Radman, M., Fons, M. & Taddei, F. 2001 Costs and benefits of high mutation rates: adaptive evolution of bacteria in the mouse gut. *Science (80-. ).* **291**, 2606–8. (doi:10.1126/science.1056421)

62. Loh, E., Salk, J. J. & Loeb, L. A. 2010 Optimization of DNA polymerase mutation rates during bacterial evolution. *Proc. Natl. Acad. Sci.* **107**, 1154–9. (doi:10.1073/pnas.0912451107)

# 7. Appendices

## Appendix A: Appearance of a double mutant

In the following analysis we assume that $0 < \mu \ll U \ll s \ll 1$ and $s/\mu < N < (s/\mu)^2$ – see model overview for details. This also means that $U + 2\mu \approx U$ and $U/s + U \approx U/s$.

The probability $q$ that a random offspring in the next generation is $AB$ given there are no $AB$ in the current generation can be approximated by:

$$q = \mu^2 e^{-\frac{U}{s}-U} + 2\frac{\mu^2}{s}e^{-\frac{U}{s}-U} = \frac{\mu^2}{s}e^{-U-\frac{U}{s}}(s+2) \approx \frac{\mu^2}{s}\left(1 - U - \frac{U}{s}\right)(2+s).$$

Using the above assumptions, this resolves to:

$$q \approx 2\frac{\mu^2}{s}\left(1 - \frac{U}{s}\right).$$

Taking the derivative with respect to $U$ and denoting $g = U/\mu$ ($g$ can be thought of as the number of non-specific loci in the genome):

$$\frac{dq}{dU} = \frac{2U(2s-3U)}{g^2 s^2} > 0 \Leftrightarrow U < \frac{2}{3}s.$$

So $q$ increases with $U$ because the right hand side is guaranteed to be true under the assumption $U \ll s$.

For a population with SIM

$$q_{SIM} = \mu^2 e^{-\frac{U}{s}-U} + 2\frac{\tau\mu^2}{s}e^{-\frac{U}{s}-\tau U} \approx \frac{\mu^2}{s}\left(1-\frac{U}{s}\right)\left(s(1-U) + 2\tau(1-\tau U)\right) = \frac{\mu^2}{s}\left(1 - \frac{U}{s}\right)(s(1-U) + 2\tau(1-U) - 2\tau(\tau-1)U) = \frac{\mu^2}{s}\left(1-\frac{U}{s}\right)\left((s+2\tau)(1-U) - 2\tau(\tau-1)U\right) \approx \frac{\mu^2}{s}\left(1-\frac{U}{s}\right)(2\tau(1-U) - 2\tau(\tau-1)U).$$

The last approximation assumes that $\tau \geq 1 \Rightarrow s \ll 2\tau$. Rearranging the last result, we find the approximation

$$q_{SIM} \approx 2\tau \frac{\mu^2}{s}\left(1 - \frac{U}{s}\right)(1 - \tau U) = \tau(1 - \tau U)q.$$

Taking the derivative with respect to $\tau$,

$$\frac{dq_{SIM}}{d\tau} = q(1 - 2\tau U) > 0 \Leftrightarrow \tau U < \frac{1}{2},$$

because $q$, $U$, and $\tau$ are all positive. So the condition $\tau U \ll s \ll 1$ guarantees that $q_{SIM}$ increases with $\tau$, and it is also sufficient for $q_{SIM} > q$ (not shown).

## Appendix B: Fixation of a double mutant

Following Eshel [50], the fixation probability $\rho$ of the double mutant $AB$ is

$$\rho = 2\frac{\alpha - 1}{\alpha} + o(\alpha - 1),$$

where $\alpha$ is the fitness of the double mutant relative to the population mean fitness $\bar{\omega}$ and assuming that fitness is measured by the average number of progeny which is Poisson distributed:

$$\alpha = \frac{(1 + sH)e^{-U}}{\bar{\omega}}.$$

Here, we only consider progeny without new deleterious mutations; their fraction is $e^{-U}$. This factor cannot be ignored because there is variation in mutation rates within the population (see "minor technical point" by Johnson and Barton [81]). At this stage, double mutants are still very rare, so we can use the population mean fitness at the MSB. The population mean fitness can be approximated by $\bar{\omega} = e^{-U}$ (see *supporting information*). Therefore,

$$\rho = 2\frac{sH}{1+sH} + o(sH).$$

Assuming $sH$ is small ($sH \ll 1$) we can approximate this by

$$\rho \approx 2sH.$$

## Appendix C: Fixation of a double mutant with SIM$_e$

With SIM$_e$ the mutation rate of $ab$ is $\tau U$ while that of $AB$ is only $U$. We assume the population reached a MSB before the fixation of $AB$ because convergence to MSB [42] is much faster than adaptation. Following the derivation in Appendix B, the relative fitness of SIM$_e$ is

$$\alpha_{SIMe} = \frac{(1+sH)e^{-U}}{e^{-\tau U}} = (1+sH)e^{U(\tau-1)}.$$

Plugging that in the fixation probability,

$$\rho_{SIM_e} \approx 2\frac{(1+sH)e^{U(\tau-1)}-1}{(1+sH)e^{U(\tau-1)}} = 2\frac{1+sH-e^{-U(\tau-1)}}{1+sH} = \rho + 2\frac{1-e^{-U(\tau-1)}}{1+sH}.$$

This can be simplified by a 1$^{st}$ order approximation for $e^{-U(\tau-1)}$:

$$\rho_{SIM_e} \approx \rho + 2\frac{U(\tau-1)}{1+sH} = \rho\left(1 + \frac{U(\tau-1)}{sH}\right).$$

Because $\frac{U(\tau-1)}{sH} > 0$, the right hand side is greater than 1. Therefore,

$$\rho_{SIM_e} > \rho.$$

Because the appearance with SIM$_e$ is the same as with CM, the adaptation rate with SIM$_e$ can be written as

$$v_{SIM_e} = Nq\rho_{SIM_e} = Nq\rho\left(1 + \frac{U(\tau-1)}{sH}\right) = v_{CM} \cdot \left(1 + \frac{U(\tau-1)}{sH}\right).$$

# Appendix D: Mean fitness at the mutation-selection balance

Denote the frequency of individuals with $x$ deleterious alleles by $f_x$. The frequency of such individuals in the next generation $f'_x$ is given by

$$\bar{\omega} f'_x = \sum_{y \geq 0} f_y m_{x,y},$$

where $m_{x,y}$ is the transition probability from $y$ deleterious alleles to $x$ deleterious alleles and $\bar{\omega}$ is the population mean fitness.

The term $m_{x,y}$ consist of the fitness of individuals with $y$ deleterious alleles, $\omega_y$, and the probability that the precise number of mutations occurred. Specifically, if $y \geq x$ then exactly $y$-$x$ beneficial mutation must occur; if $y \leq x$ then exactly $x$-$y$ deleterious mutations must occur:

$$m_{x,y} = \begin{cases} \omega_y \cdot P(x - y \text{ deleterious mutations}), & y < x \\ \omega_y \cdot P(y - x \text{ beneficial mutations}), & y > x \\ \omega_y \cdot P(no\ mutations), & y = x \end{cases}.$$

Using the probability mass function of a Poisson distribution, we can expand the above equation to

$$\bar{\omega} f'_x = \sum_{y \leq x} f_y \omega_y \frac{e^{-\delta U_y}(\delta U_y)^{x-y}}{(x-y)!} + \sum_{y \geq x} f_y \omega_y \frac{e^{-\beta U_y}(\beta U_y)^{y-x}}{(y-x)!}, \quad \forall x \geq 0,$$

where $\omega_y$ is the fitness of individuals with $y$ deleterious alleles, $\bar{\omega}$ is the population mean fitness ($\bar{\omega} = \sum \omega_x f_x$), $\delta$ and $\beta$ are the fraction of mutations that are deleterious and beneficial, respectively ($\delta + \beta = 1$ and $0 \leq \beta < \delta \leq 1$), and $U_y$ is the average number of new mutations per generation in an individual with $y$ deleterious alleles.

This can be written as a matrix equation by multiplying the frequencies vector $f$ and the mutation-selection matrix $M$:

$$\bar{\omega} f' = Mf$$

At the mutation-selection balance (MSB), $f^*$ solves the equation (a star * denotes equilibrium quantities)

$$\bar{\omega}^* f^* = Mf^*.$$

Without beneficial mutations ($\delta=1$ and $\beta=0$), the above equation simplifies to

$$\bar{\omega} f'_x = \sum_{y \leq x} f_y \omega_y \frac{e^{-U_y} U_k^{x-y}}{(x-y)!}, \quad \forall x \geq 0$$

and $M$ is a triangular matrix. In this case the population mean fitness can be found by solving the equation for $f_0$:

$$\bar{\omega} f_0 = f_0 m_{0,0} = f_0 \omega_0 e^{-U_0} \Rightarrow \bar{\omega} = \omega_0 e^{-U_0},$$

which means that the population mean fitness is equal to the product of the fitness of mutation-free individuals and the probability that a mutation-free individual does not mutate. If $\omega_x = (1-s)^x$ and $U_x = U$ (constant uniform mutation rate) then $\bar{\omega} = e^{-U}$ [82] and by the forward substitution method the frequencies vector is

$$f_x = e^{-U/s} (U/s)^x / x!,$$

that is, the number of deleterious mutations per individual is Poisson distributed with average $U/s$ [40]. With constitutive mutagenesis (CM), the population mean fitness at the MSB is $e^{-\tau U}$: it decays exponentially as a function of $\tau$ the mutation rate fold increase. In contrast, stress-induced mutagenesis (SIM), as shown by Agrawal [35], does not change the population mean fitness with respect to normal mutagenesis (NM). This is because the least loaded individuals ($x=0$), with fitness $\omega_0$=1, also have the lowest mutation rate, *U*, and therefore the population mean fitness is $e^{-U}$.

With beneficial mutations (β>0), the matrix *M* is a positive matrix, and by the *Perron-Frobenius Theorem* (Otto and Day 2007, p. 709) $\overline{\omega}^*$ is the largest eigenvalue of *M* and *f\** is its unique positive eigenvector with $\sum f = 1$.

This eigenvalue problem is hard to solve analytically, however, by neglecting elements outside the main three diagonals of *M* we have shown before [37] that the population mean fitness increases with the mutation rate of individuals with a below–average fitness:

$$sign \frac{\partial \overline{\omega}}{\partial U_x} = sign(\overline{\omega} - \omega_x).$$

Nevertheless, this framework allows the numerical calculation of the population mean fitness for finite *n*-by-*n* mutation-selection matrices by defining *n* such that $\omega_x = 0 \ \forall x \geq n$. The mean fitness of populations with different mutational strategies is then calculated by manipulating $U_x$.

Figure D1 shows that $e^{-U}$ is a good approximation to the population mean fitness (because $\beta \ll \delta < 1$), and that SIM slightly increases the population mean fitness with respect to NM; a sufficient condition is that the mutation rate of individuals with below average fitness is increased [37]. Since we assume that $U < s$, then $e^{-U} \approx 1 - U > 1 - s$. Therefore, for SIM to increase the population mean fitness it must increase the mutation rate in individuals with at least one deleterious mutation.

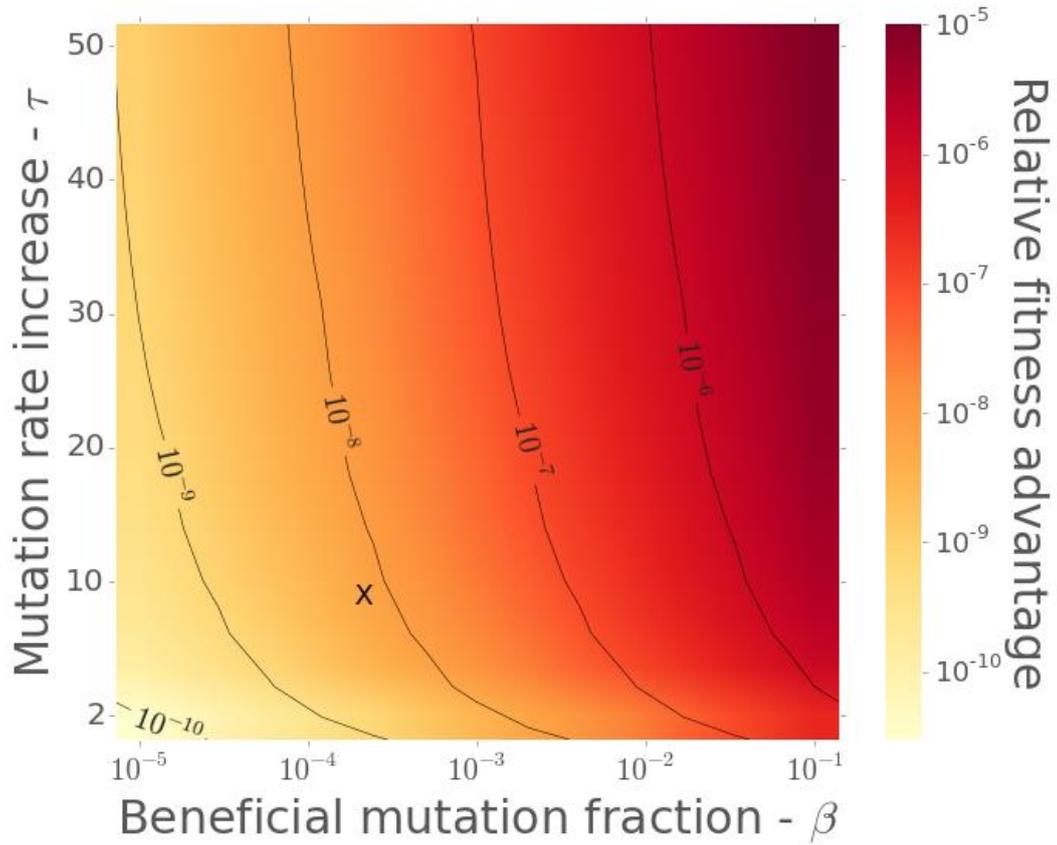

**Figure D1 – Mean fitness at the mutation-selection balance with stress-induced mutagenesis.** The brightness represents the fitness advantage of stress-induced mutagenesis over normal mutagenesis (($\bar{\omega}_{SIM} - \bar{\omega}_{NM})/\bar{\omega}_{NM}$) at the mutation-selection balance. The x-axis is the fraction of mutations that are beneficial $\beta$. The y-axis is the mutation rate fold increase under stress $\tau$. "X" marks the parameter set $\beta$=1/5000 and $\tau$=10, in which the fitness advantage of SIM is ~5·10$^{-9}$.

# Appendix E: Possible relationships between stress and mutation

In the main text we used a threshold relationship between stress and mutation: if fitness drops below a threshold (<1 for SIM, ≤1 for $SIM_e$), the mutation rate increases $\tau$-fold. But the relationship between stress and mutation can be more complex. For example, Agrawal [35] has used a continuous relationship defined by a curvature parameter $k$. This relationship defines the mutation rate for an individual with fitness $\omega$, baseline mutation rate $U$, and a maximal mutation rate fold increase $\tau$ as

$$U(\omega) = \tau U - (\tau - 1)U\omega^k.$$

When $k$ approaches 0 this expression approaches $U$, corresponding to the NM strategy. When $k$ approaches infinity this expression approaches eq. 1(1), corresponding to the SIM threshold strategy. See Figure E1 for a plot of these continuous relationships for various values of $k$.

Figure E2 shows the adaptation time for three continuous strategies ($k$=1/10, 1, and 10). Remarkably, the dynamics of a continuous strategy can be approximated by a threshold strategy by matching the mutation rates of single mutants:

$$U(\omega) = \begin{cases} U, & \omega \geq 1 \\ \tau U - (\tau - 1)U(1-s)^k, & \omega < 1 \end{cases}.$$

This is equivalent to using a threshold strategy with mutation rate increase $\tau-(\tau-1)(1-s)^k$. The dashed lines in Figure E2 demonstrate this approximation. The continuous strategies can be approximated by threshold strategies because the main factor determining the adaptation rate is the mutation rate increase of the wildtype and the single mutants (*ab*, *aB*, and *Ab*). This is because individuals with more than a single mutation do not have a significant contribution to adaptation.

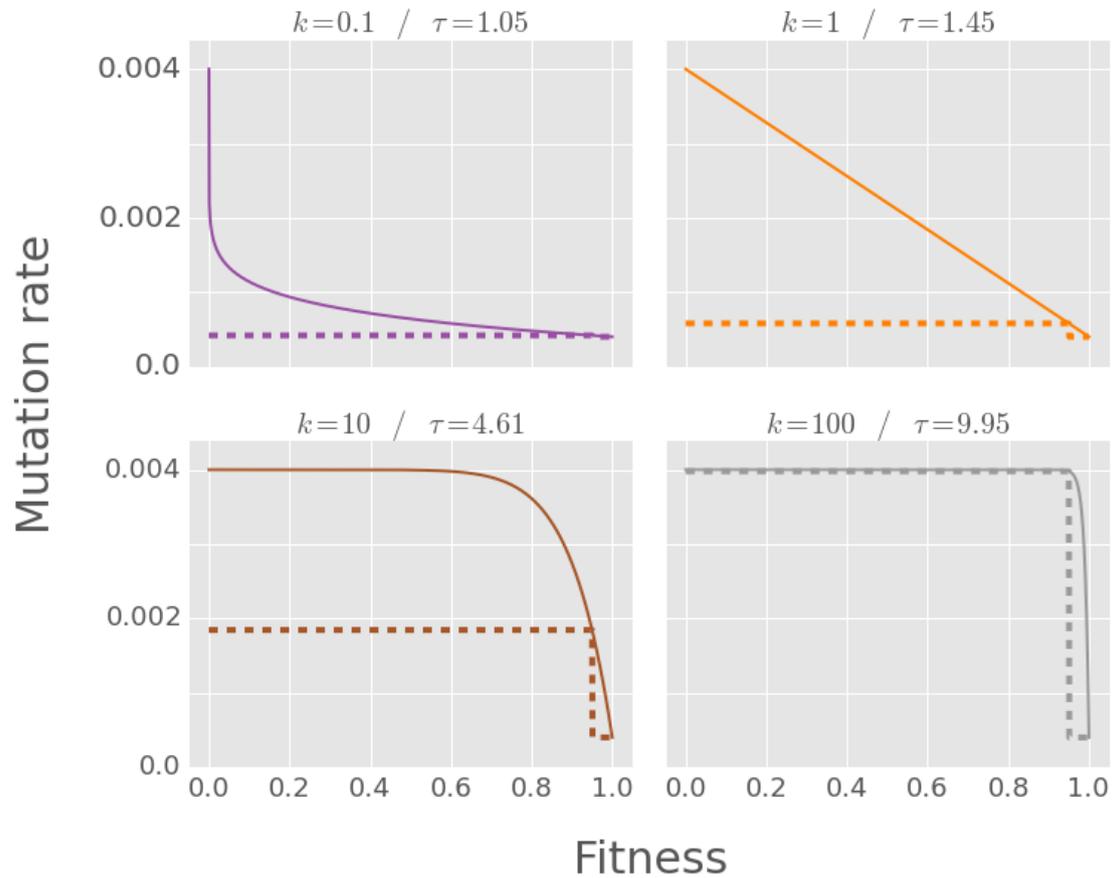

**Figure E1– Different relationships between stress and mutation.** The figure shows continuous relationships between fitness (x-axis) and mutation rate (y-axis) in solid lines and threshold relationships in dashed lines. The threshold relationship is defined in section 2 of the main text. The continuous relationships are defined in Supporting Text S4. Each panel shows a pair of relationships, with *k* increasing from 1/10 (convex relationship), to 1 (linear relationship) to 10 and 100 (concave relationships). Each continuous relationship is compared with a threshold relationship that has the same mutation rate for wildtypes (*ab/0*) and single mutants (*Ab/0*, *aB/0*, *ab/1*). Figure S5 shows that the adaptation rate with such threshold relationship approximates the adaptation rate with a continuous relationship.

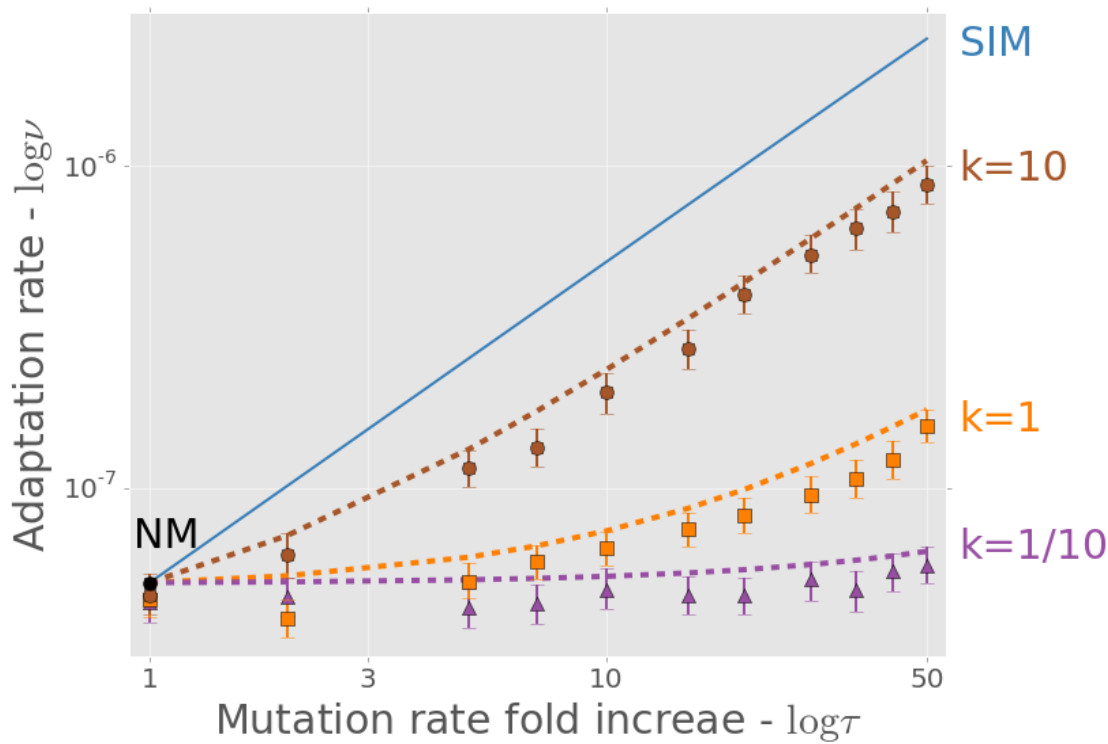

**Figure E2 – Complex adaptation with continuous relationship stress-induced mutagenesis.** The figure shows the adaptation rate $v$ as a function of the mutation rate increase $\tau$ (both in log scale). The solid line is an analytical approximation of SIM (same as in Figure 2). Markers are the results of simulations of adaptation with SIM with different continuous relationships between fitness and mutation rate. These continuous relationships are defined by a mutation rate fold increase $\tau$=10 and a curvature parameter $k$=10, 1, and 1/10, top to bottom (see Supporting Text S4 and Figure S3 for more details on continuous SIM). Each dashed line is an approximation of a continuous SIM using a SIM threshold strategy (eq. 13**Error! Reference source not found.**) with $\tau$=4.61, 1.45, and 1.05, top to bottom. The fit between the dashed lines and the corresponding markers suggests that a threshold strategy captures the adaptive dynamics. Error bars represent 95% confidence interval of the mean (at least 1,000 simulations per point; computed with bootstrap with 1,000 samples per point). Parameters (see Table 1): $U$=0.0004, $s$=0.05, $\beta$=0.0002, $H$=2, $N$=$10^6$.

# Appendix F: Competitions between mutational strategies

We also simulated direct competitions between the different mutational strategies (NM, CM, and SIM). In these competitions, half of the population alters its mutational strategy to an invading strategy at the time of the environmental change. Each simulation provides a sample of the frequency of the invading strategy after the appearance and subsequent fixation or extinction of the double mutant *AB*. If the average final frequency is significantly lower or higher than 50% we consider the invading strategy disfavored or favored by natural selection over the initial strategy. Statistical significance was calculated using a 1-sample 2-tailed t-test.

Figure F1 summarizes the competitions. CM clearly loses to both SIM and NM (first and second panels from the right). SIM is significantly advantageous over NM when the mutation rate increase is large enough ($\tau>2$; 2-tail t-test, $P<0.0015$).

These results show that the evolutionary advantage of SIM at the population-level corresponds to an individual-level advantage and can lead to the evolution of stress-induced mutagenesis by natural selection, even when constitutive mutagenesis is strongly disfavored. This is consistent with previous results in smooth fitness landscapes [37].

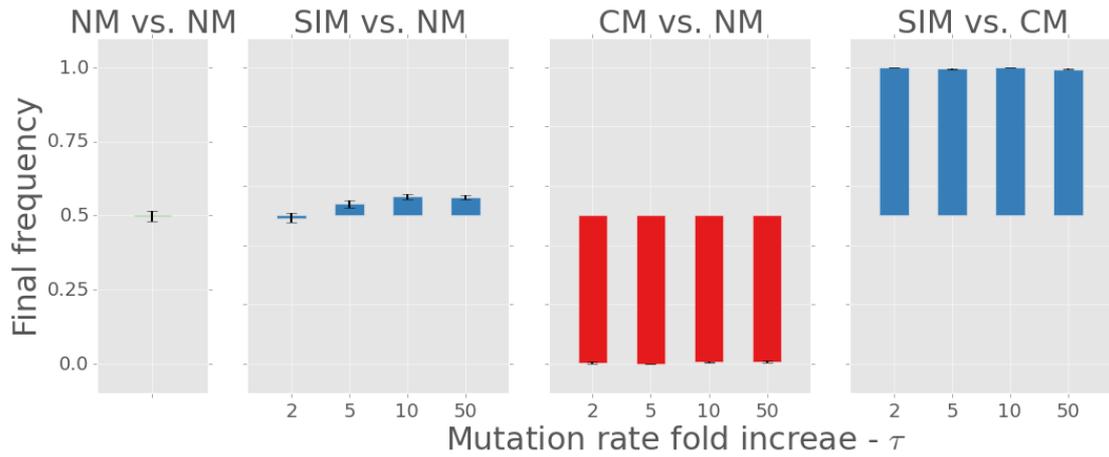

**Figure F1 – Direct competitions between three mutational strategies.** The figure shows the average final frequency of (from right to left): stress-induced mutagenesis (SIM) vs. constitutive mutagenesis (CM); CM vs. normal mutagenesis (NM); SIM vs. NM; and NM vs. NM (control). Initial frequencies are always 0.5. Several mutation rate fold increases are shown on the x-axis. SIM defeats CM and is significantly advantageous over NM when τ>2 (2-tail t-test, P<0.0015). CM losses to NM and SIM (P≈0). Therefore, SIM is favored by selection over both NM and CM. Changing roles between resident and invader didn't affect the results (not shown). Error bars represent the standard error of the mean (500 simulations per point). Parameters (see Table 1): U=0.0004, s=0.05, β=0.0002, H=2, N=10^6.

## 8. Supporting Figures

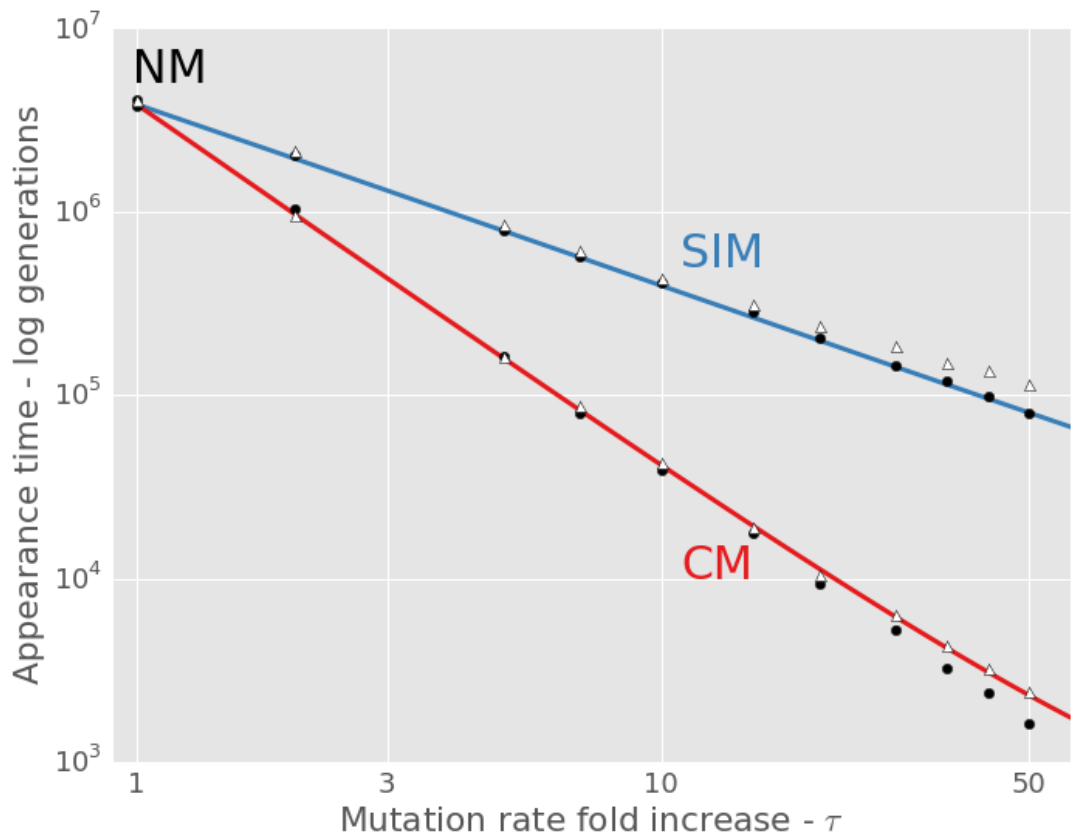

**Figure S1 – Waiting time for the appearance of a double mutant** as a function of the mutation rate fold increase $\tau$. Normal mutagenesis (NM) is $\tau=1$; constitutive mutagenesis (CM) in red; stress-induced mutagenesis (SIM) in blue. Lines are analytic approximations (eqs. 2, 3 in main text). Markers are means of simulation results - black circles for the standard simulations, white triangles for alternative simulations in which *AB* cannot appear on deleterious backgrounds. The standard error of the mean was too small to show. At least 1,000 simulations per point. Both axes are in log scale. The appearance time decreases as a function of $\tau^2$ and $\tau$ with CM and SIM, respectively. Appearance time is slightly longer if *AB* only appears on unloaded background (white triangles) which explains the difference between the analytic approximations and the simulation results for SIM in Figure 2. Parameters are the same as in Figure 2.

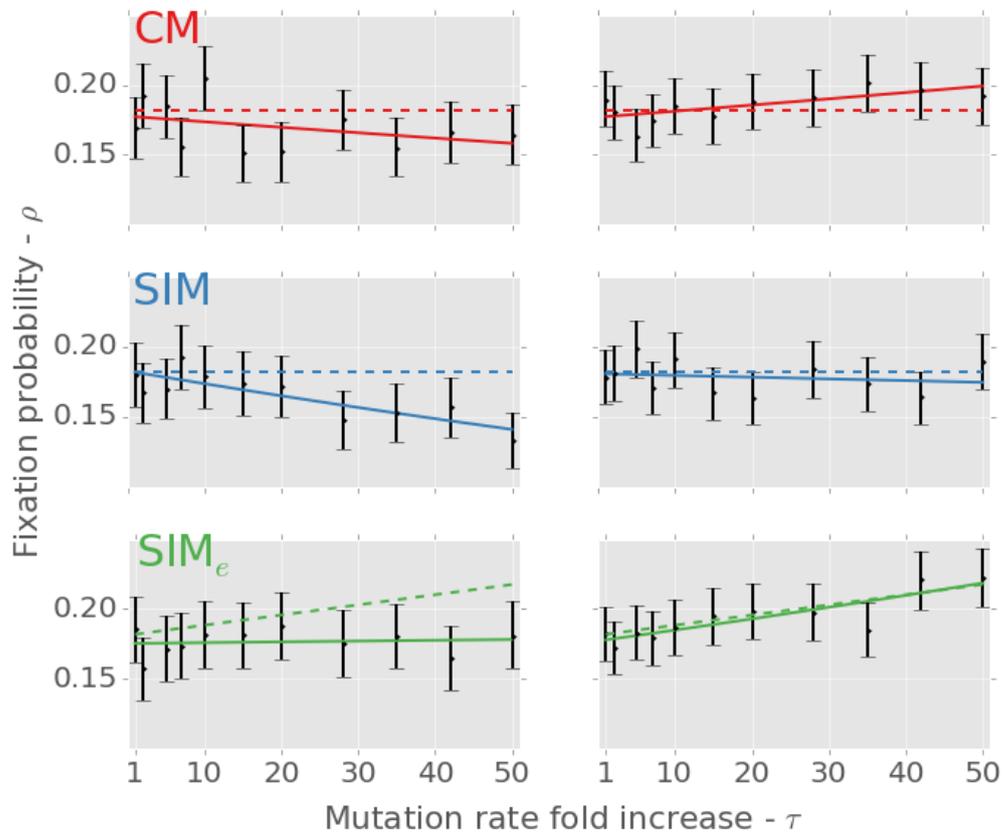

**Figure S2 – Fixation probability of the double mutant *AB*** as a function of the mutation rate fold increase $\tau$ with three mutational strategies: constitutive mutagenesis (CM; top panels in red), stress-induced mutagenesis (SIM; middle panels in blue) and stress-induced mutagenesis with environmental stress (SIM$_e$; bottom panels in green; see section 3.5 in main text). Dashed lines are analytic approximations; black error bars represent simulation results with 95% confidence interval of the mean (at least 1,000 simulations per point; computed with bootstrap with 10,000 samples per point); solid lines are the logistic regression lines computed from the simulation results. The three left panels are results of the standard simulations. The three right panels are results of simulations in which *AB* cannot appear on deleterious backgrounds - in these cases there is no significant difference between the simulation results and our analytic approximations (compare solid and dashed lines; regression slope tests with $\alpha$=0.05). However, if *AB* can appear on a deleterious background (left panels) then its fixation probability is lower [81]. For example, the fixation probability of *AB* with a single deleterious mutation is $\rho_{AB/1} = \rho \left(1 - \frac{1}{H(1-s)}\right) < \rho$. In addition, the figure shows that SIM$_e$ has a higher fixation probability than CM and SIM: the green lines, representing SIM$_e$, are always higher than the red and blue lines representing CM and SIM. Parameters are the same as in Figure 2.